\documentclass[sigconf]{acmart}

\usepackage{booktabs} % For formal tables

\hypersetup{draft}
\usepackage{notoccite}
\renewcommand\footnotetextcopyrightpermission[1]{} % removes footnote with conference information in first column
\begin{document}
\title{Heterogeneous FPGA+GPU Embedded Systems: Challenges and Opportunities }

\author{Mohammad Hosseinabady}
\orcid{1234-5678-9012}
\affiliation{%
  \institution{University of Bristol}
  \city{Bristol, UK}
}
\email{m.hosseinabady@bristol.ac.uk}

\author{Mohd Amiruddin Bin Zainol}

\affiliation{%
	\institution{University of Bristol}
	\city{Bristol, UK}
}
\email{mb14650@bristol.ac.uk}

\author{Jose Nunez-Yanez}

\affiliation{%
	\institution{University of Bristol}
	\city{Bristol, UK}
}
\email{J.L.Nunez-Yanez@bristol.ac.uk}

\begin{abstract}
The edge computing paradigm has emerged to handle cloud computing issues such as scalability, security and high response time among others. This new computing trend heavily relies on ubiquitous embedded systems on the edge. Performance and energy consumption are two main factors that should be considered during the design of such systems. Focusing on performance and energy consumption, this paper studies the opportunities and challenges that a heterogeneous embedded system consisting of embedded FPGAs and GPUs (as accelerators) can provide for applications. We study three \textit{design}, \textit{modeling} and \textit{scheduling} challenges throughout the paper. We also propose three techniques to cope with these three challenges. Applying the proposed techniques to three applications including image histogram, dense matrix-vector multiplication and sparse matrix-vector multiplications show  
1.79x and 2.29x improvements in performance and energy consumption, respectively, when both FPGA and GPU execute the corresponding application in parallel.
 
\end{abstract}


%\ccsdesc[500]{Computer systems organization~Embedded systems}

%\keywords{ACM proceedings, \LaTeX, text tagging}

\maketitle

\section{Introduction}
\label{sec:Introduction}

The emergence of edge computing, which brings the analytics, decision making, automation and security tasks close to the source of data and applications, has raised new opportunities and challenges in the area of IoT and embedded systems. This new computing trend enables the execution of cloud-native tasks on resource-limited embedded systems. The versatile and dynamic behavior of these tasks has changed the traditional definition of an embedded system that has been mainly defined as a small system tuned to efficiently run a specific task inside a big system. Recently Google has introduced the tensor processing unit (TPU) to efficiently run neural-network-based machine learning algorithms on the edge \cite{Jouppi:2017:IPA:3079856.3080246}. Amazon has announced the AWS Greengrass to bring cloud computing to the edge \cite{amazon-aws-greengrass}.

New embedded systems demand new features such as efficiently working with Internet, enabling highly computational power, consuming low energy, providing real-time at the scale of machinery with nanosecond latency and working collaboratively with other similar systems to finish a shared task.  Heterogeneous embedded systems are promising techniques to cope with these ever-increasing demands. Toward this end, FPGAs and GPUs, the two common accelerators, have separately been integrated into embedded systems recently, by industry, to address the new requirements.  However, integrating them in an embedded system to collaboratively execute a complex task, fulfilling the performance, latency, predictability, and energy consumption constraints, is still a challenge.

\begin{figure}
	\centering
	\includegraphics[width=1\linewidth]{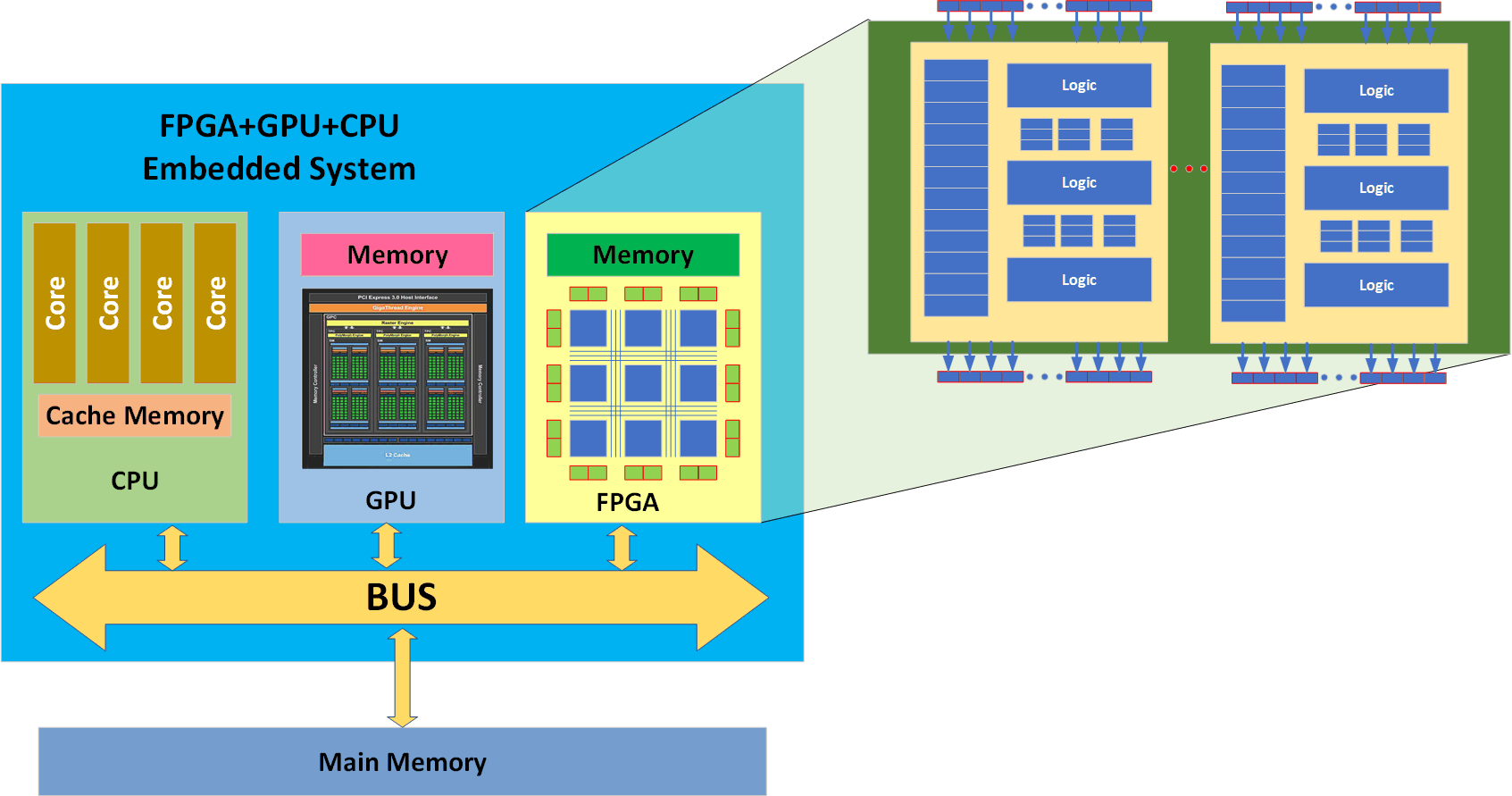}
	\caption{Embedded FPGA and GPU in a System}
	\label{fig:EmbeddedFPGAandGPUinaSystem}
\end{figure}

Fig.~\ref{fig:EmbeddedFPGAandGPUinaSystem} shows the overview of an embedded system consisting of three processing elements (PEs) including a multi-core CPU, a many-core GPU and an FPGA. The main feature of this architecture is the direct access of PEs to the main memory using the same address space and shared memory controller, in contrast to the current desktop platforms with FPGAs and GPUs that communicate via PCIe with system memory. This feature enables the accelerators to benefit from \textit{zero-copy} data transfer technique without the performance and energy overhead of the PCIe in between, which improves the memory bandwidth utilization and reduce the inter PEs communication overhead. Therefore, each PE can potentially achieve its high performance in executing an application. However, choosing a proper PE to run a given task, with maximum performance and minimum energy consumption, is not an easy decision to make. To make this process clear, we study and compare the performance and energy consumption of accelerators (i.e. the GPU and FPGA), running different tasks.

To this end, we need a programming model for each PE, considering the type of applications. There are many academic and industrial programming models, libraries and tools to efficiently implement different applications on embedded CPUs and GPUs. However, there is no specific design methodology for using embedded FPGAs in a system in spite of available high-level synthesis (HLS) tools based on C/C++, SystemC and OpenCL languages. This is mainly because, in the FPGA-based accelerator design, designers should first provide a hardware architecture suitable for a given task and then implement the task algorithm, accordingly. This process makes the FPGA-based accelerator design complex which needs more research to find systematic approaches for addressing different types of applications.  

In summary, three main challenges in designing a heterogeneous FPGA+GPU platform should be studied, which are as follows.
\begin{itemize}
\item \textit{Design challenge}: implementing a given task on FPGA that can compete with that of the GPU
\item \textit{Modeling challenge}: evaluating and predicting the performance and energy consumption of FPGA and GPU
\item \textit{Scheduling challenge}: distributing parallel task between FPGA and GPU in order to optimize the overall performance and energy consumption
\end{itemize}

Focusing on embedded FPGA and GPU, this paper explains the opportunities that addressing the above challenges can bring to the edge computing platforms.  
We, first, propose a systematic stream computing approach for implementing various applications on embedded FPGAs using HLS tools and then study the opportunities and challenges that a synergy among FPGA and GPU in an embedded system can provide for designers. We study a few applications that their collaborative execution on the heterogeneous system brings higher performance and lower energy consumption. We show that the collaboration between embedded FPGA and GPU can bring a renaissance to the edge computing scenario.

The rest of this paper is organized as follows. The next section explains the motivations and  contributions behind this paper. The previous work is reviewed in Section~\ref{sec:Previouswork}. The proposed FPGA stream computing engine is discussed in Section~\ref{sec:DesignChallenge}. Section~\ref{sec:ModellingChallenge} studies the performance and power modeling techniques. The scheduling challenge is explained in Section~\ref{sec:SchedulingChallenge}. The experimental setup is addressed in Section~\ref{sec:ExperimentalSetups}. Section~\ref{sec:ExperimentalResults} explains the experimental results. Finally, Section~\ref{sec:Conclusions} concludes the paper.

\section{Motivations and Contributions}
\label{sec:MotivationsandContributions}
\begin{figure}
	\centering
	\includegraphics[width=1\linewidth]{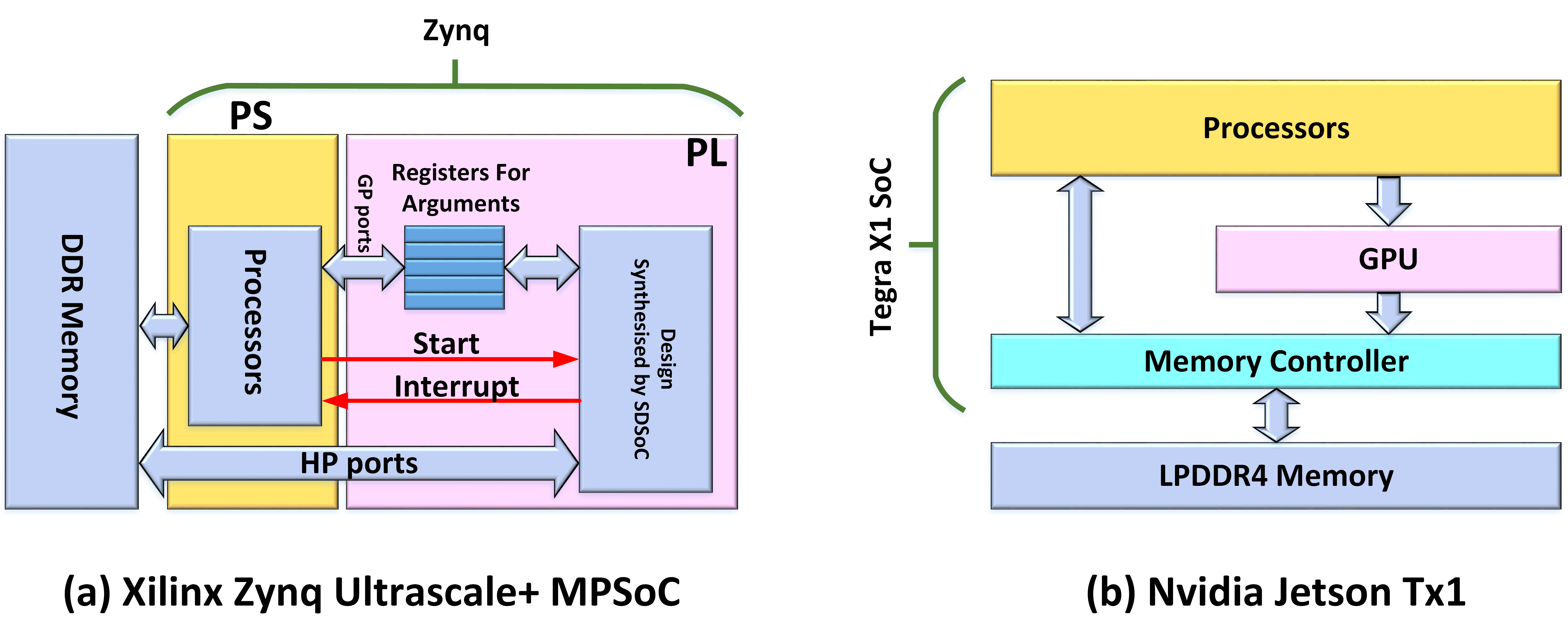}
	\caption{FPGA and GPU Embedded Systems}
	\label{fig:FPGAandGPUEmbeddedSystems}
\end{figure}

Taking the histogram operation, one of the common tasks in image processing, data mining, and big-data analysis, this section explains the motivations and contributions behind this paper. For this purpose, we have considered two embedded systems including Nvidia Jetson TX1~\cite{Nvidia-JetsonTX1} and Xilinx Zynq MPSoC (ZCU102 evaluation board)~\cite{XilinxZynqMPSOC-TRM}. Fig.~\ref{fig:FPGAandGPUEmbeddedSystems} shows the block diagrams of different parts in these systems. The Zynq MPSoC, in Fig.~\ref{fig:FPGAandGPUEmbeddedSystems}(a),  mainly consists of two parts: Processing System (PS) and Programmable Logic (PL). These two subsystems have a direct access to the system DDR memory. The PL (i.e., FPGA) performs its memory transaction through a few high-performance ports including four HPs, two HPCs, and an ACP ports. In this paper, we focus on four HP ports that can collaboratively transfer data between FPGA and memory, utilizing all the memory bandwidth available to the FPGA. 
The Nvidia Jetson TX1, shown in Fig.~\ref{fig:FPGAandGPUEmbeddedSystems}(b), is a system-on-module (SoM) combining the Nvidia Tegra X1 SoC with 4GB LPDDR4 memory and some other modules~\cite{Nvidia-JetsonTX1}. The Nvidia Tegra X1 SoC consists of a Maxwell GPU with 256 CUDA cores, 1.6GHz/s, 128K L2 cache, and 4 channel x 16bit interface to access the system memory.

Two efficient implementations of the histogram are provided for the two embedded systems. The CUDA language is used for the GPU implementation in which the NVIDIA Performance Primitives (NPP) library~\cite{Nvidia-NPP} is used. In addition, the C++ language and the Xilinx SDSoC toolset are used for the FPGA implementation which is based on the streaming pipelined computing approach similar to \cite{37325cedcd8d4056bcd3c16fc9b552d9}. This implementation reads data from the system memory and modifies the histogram bins in each clock cycle.

\begin{figure*}
	\centering
	\includegraphics[width=0.8\linewidth]{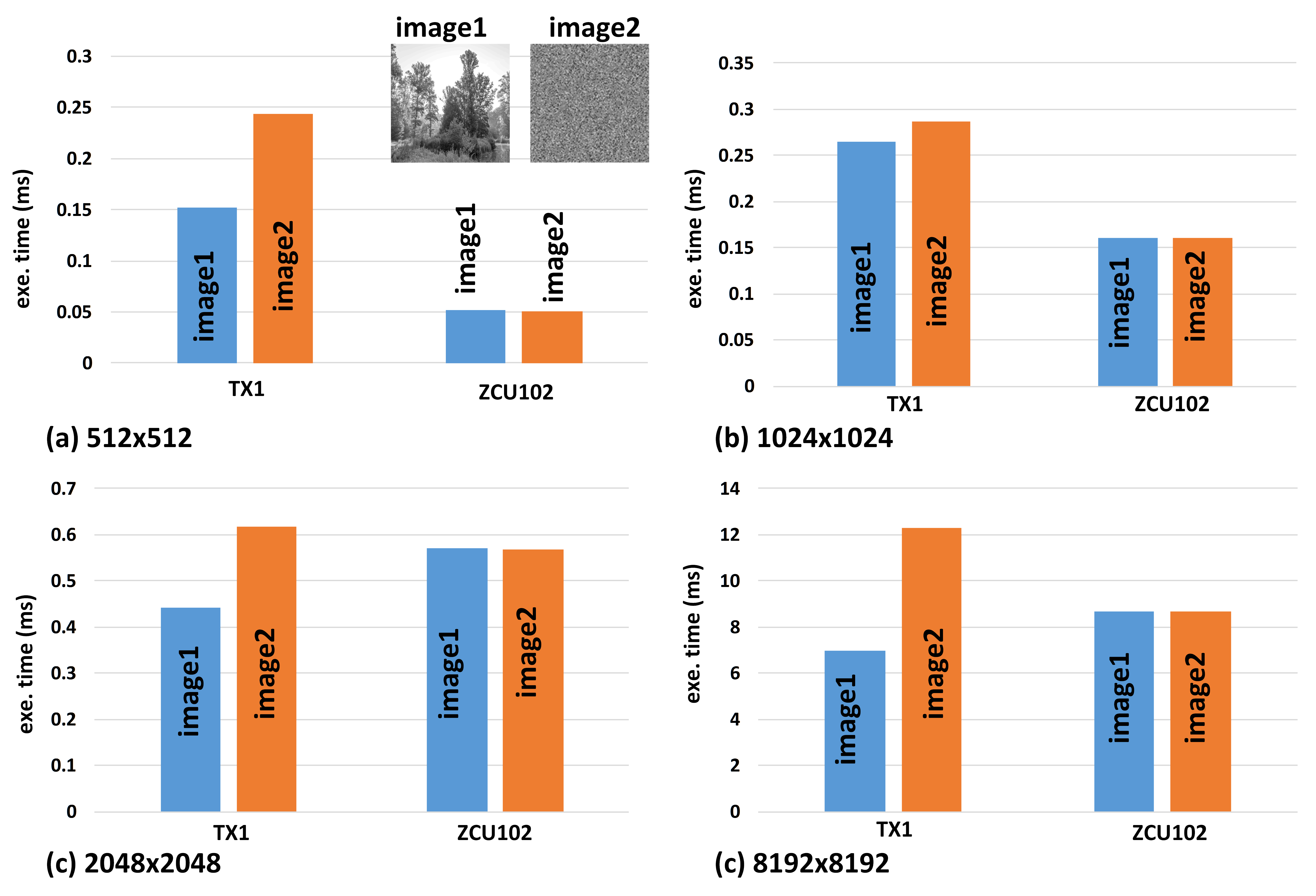}
	\caption{Histogram execution time on Jetson TX1 and Xilinx ZCU102 for two different images}
	\label{fig:HistogramexecutiontimeonJetsonTX1andXilinxZCU102fortwodifferentimages}
\end{figure*}

Fig.~\ref{fig:HistogramexecutiontimeonJetsonTX1andXilinxZCU102fortwodifferentimages} shows the execution time of the histogram operator running on the two different embedded systems considering two separate images, denoted by \textit{image1} and \textit{image2}, with different sizes ($ 512\times 512 $, $ 1024\times 1024 $, $ 2048\times 2048 $, and $ 8192\times8192 $ ). Whereas \textit{image1} is based on a real picture, \textit{image2} contains only randomly generated pixels.
As can be seen, the FPGA shows better performance in most cases and its performance does not depend on the image content, resulting in a deterministic behavior that is predictable if the image data size is known. However, the performance of the histogram implementation on the GPU depends on the image content which makes the prediction difficult even if the image size is known a priori.  Note that in two cases of \textit{image1}($ 2048\times 2048 $) and \textit{image1}($ 8192\times 8192 $) the GPU implementation is faster than that of the FPGA. 

\begin{figure}
	\centering
	\includegraphics[width=1\linewidth]{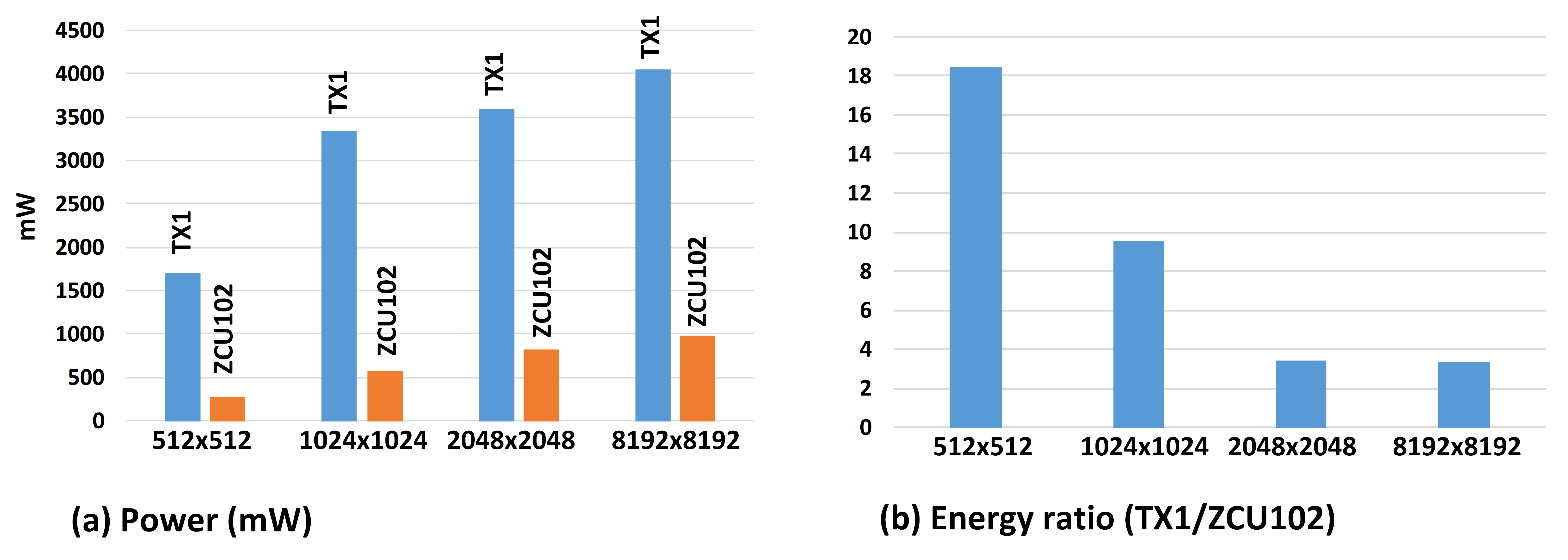}
	\caption{Histogram: Power and Energy Consumption}
	\label{fig:HistogramPowerandEnergyConsumption}
\end{figure}

Fig.~\ref{fig:HistogramPowerandEnergyConsumption} depicts the power and energy consumption of the histogram. Fig.~\ref{fig:HistogramPowerandEnergyConsumption}(a) shows the power consumption on the two embedded systems for different image sizes. As can be seen, the embedded FPGA shows much less power consumption than that of the embedded GPU.  

\begin{comment}
To increase our insight into the performance and power consumption behaviour of applications on embedded FPGA and GPU, Fig.~\ref{fig:SpMVexecutiontimeandPowerconsumption} shows the execution time and power consumption of sparse matrix vector multiplication (SpMV) on the embedded FPGA and GPU. The SpMV performance heavily depends on the sparsity and the pattern on non-zero (nnz) elements in the input matrix. The FPGA implementation of the SpMV uses the concepts of streaming pipelined computation and the GPU implementation is based on the Nvidia cuSPARSE library. Whereas Fig.~\ref{fig:SpMVexecutiontimeandPowerconsumption}(a) shows the execution time in \textit{ms}, Fig.~\ref{fig:SpMVexecutiontimeandPowerconsumption}(b) depicts the normalized power consumption to that of the FPGA implementation. These diagrams confirm that FPGA consumes less energy than GPU and can provide a performance comparable to that of the GPU for streaming applications. Note that in this example, FPGA shows better performance for small data sizes, whereas GPU is faster for large data sets.
\begin{figure}
	\centering
	\includegraphics[width=1\linewidth]{motivation-spmv-performance.png}
	\caption{SpMV: execution time and Power consumption}
	\label{fig:SpMVexecutiontimeandPowerconsumption}
\end{figure}
\end{comment}

\begin{figure}
	\centering
	\includegraphics[width=0.8\linewidth]{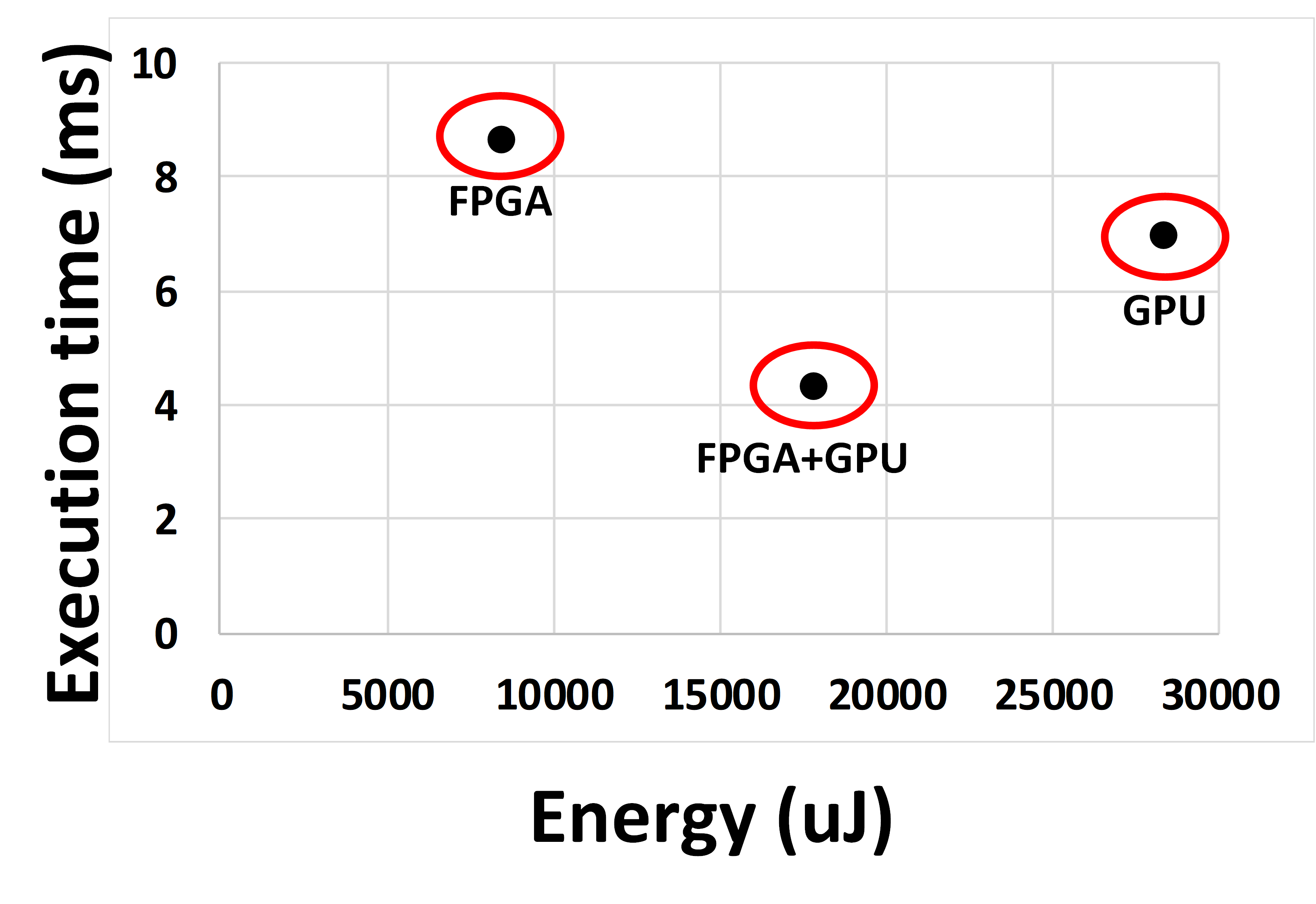}
	\caption{Histogram: Performance and Energy trade-off}
	\label{fig:Histogram:PerformanceandEnergytradeoff}
\end{figure}

Now if we equally divide the \textit{image1} of size $ 8192\times 8192 $ between the embedded FPGA and GPU, then the execution time on FPGA and GPU would be about $ 3.51 ms $ and $ 4.35 ms $, respectively which improves the performance by a factor $ 6.99/4.35= 1.6 $. In this case, the FPGA and GPU energy consumptions are $ 4133.8 \mu J$  and $ 13653.9 \mu J $, respectively which improves the total energy consumption by a factor of $ 1.59 $. Fig.~\ref{fig:Histogram:PerformanceandEnergytradeoff} shows the trade-off between the energy consumption and performance for running the histogram on FPGA, GPU and both.

This trade-off has motivated us to study the performance and energy consumption of different applications on both platforms and propose an FPGA+GPU based embedded systems to improve the total performance and energy consumption by scheduling a given task between these two accelerators. 
The main contributions of this paper are as follows:
\begin{itemize}
\item Studying the challenges of design, modeling and scheduling on FPGA+GPU embedded systems.
\item Clarifying the opportunities that addressing these challenges provide 
\item Proposing a stream computing technique on FPGA to deal with the design challenge
\item Modelling the FPGA performance and power consumption to cope with the modeling challenge.
\item Proposing an FPGA+GPU embedded system to improve performance and energy consumption to address the scheduling challenge
\end{itemize}

\section{Previous work}
\label{sec:Previouswork}
There have been extensive studies on employing GPU and FPGA on desktop and cloud servers in the literature. 

An OpenCL-based FPGA-GPU implementation for the database join operation is proposed in \cite{8124981}. They use the Xilinx OpenCL SDK (ie.e, SDAccel) to explore the design space. A real-time embedded heterogeneous GPU/FPGA system is proposed by \cite{7816978} for radar signal processing. An energy-efficient sparse matrix multiplication is proposed in \cite{7482073} which utilizes the GPU, Xeon Phi, and FPGA. An FPGA-GPU-CPU heterogeneous architecture has been considered in \cite{6412108} to implement a real-time cardiac physiological optical mapping. All these systems use the PCIe to connect the GPU and FPGA to the host CPU. In contrast to these approaches, we assume a direct connection between the accelerators and the system memory. 

A heterogeneous FPGA/GPU embedded system based on the Intel Arria 10 FPGA and the Nvidia Tegra X2 is presented in \cite{8502371} to perform ultrasound imaging tasks. In contrast to this approach, we study the challenges and opportunities that hybrid FPGA/GPU embedded systems can bring to the edge computing by considering wider types of tasks and applications.

\section{Design Challenge}
\label{sec:DesignChallenge}
This paper considers streaming applications which can receive data, perform computation, and generate results in a pipeline fashion. Many tasks can be categorized as streaming applications, among them are data parallel, window, and block processing tasks~\cite{6574848}. There are many techniques and research that show how to map a streaming application on GPUs~\cite{6574848,6131835,7559535,8035150,6510489,5437735}, however, efficiently mapping these applications on FPGAs, using a systematic approach, requires more research.

\begin{figure}
	\centering
	\includegraphics[width=1\linewidth]{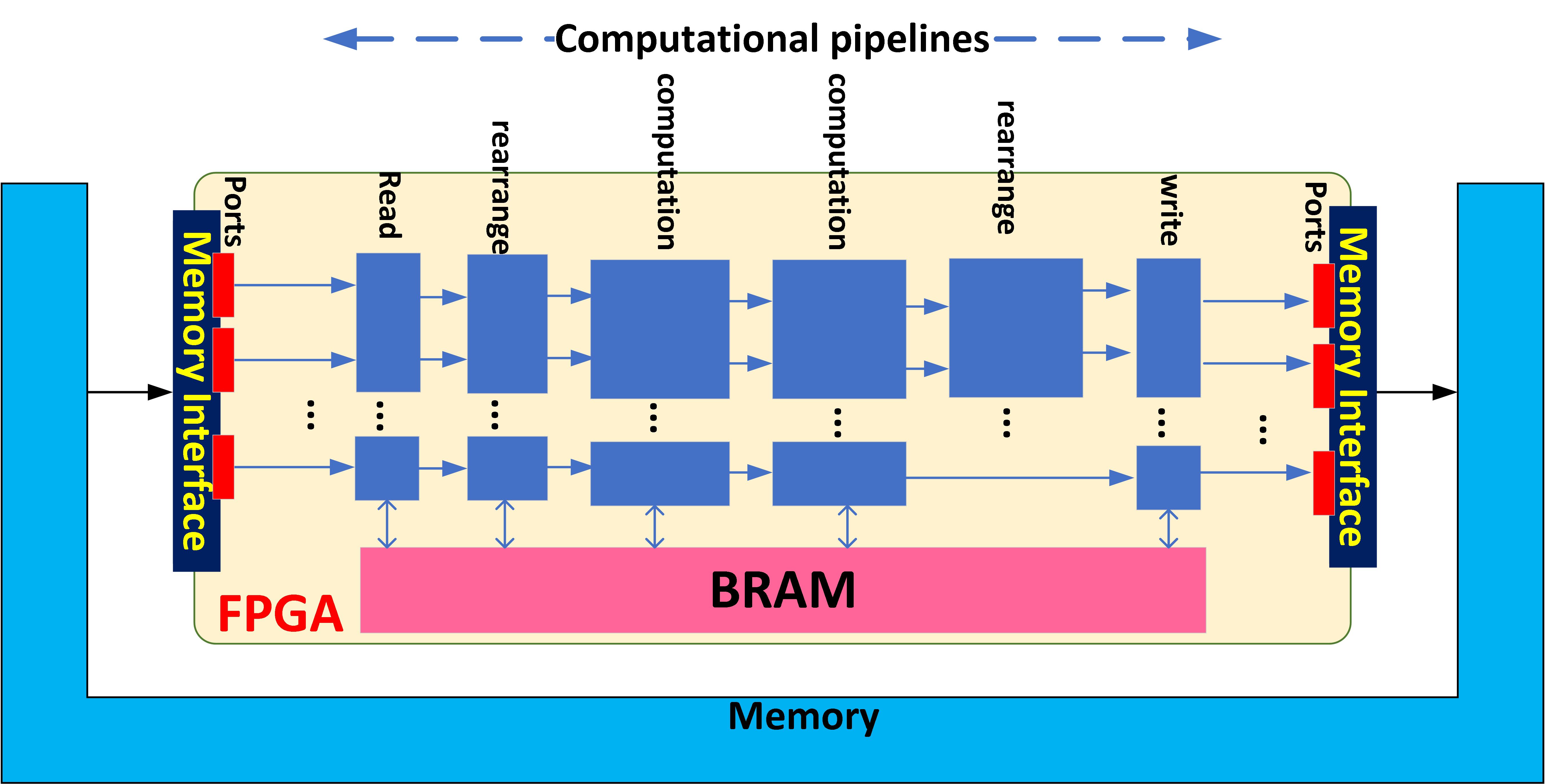}
	\caption{Overview of Stream computing engine on FPGA}
	\label{fig:OverviewofStreamcomputingengineonFPGA}
\end{figure}

Traditionally, FPGA accelerators are designed by Hardware Description Languages (HDLs) that can potentially provide a high-performance implementation. However, the HDL based design flow is tedious and time-consuming. In addition, the design is not easily adaptable (modifiable) to the versatile edge computing environment that includes a variety of algorithms with different configurations and complexity. To alleviate these issues, High-Level Synthesis (HLS) has been proposed by academia and industry that is increasingly popular for accelerating algorithms in FPGA-based embedded platforms. Studies have shown that HLS can provide high-performance and energy-efficient implementations with shortening time-to-market and addressing today's system complexity~\cite{7368920}. Following the HLS design flow, we propose a streaming pipelined computing engine to implement several applications. Fig.~\ref{fig:OverviewofStreamcomputingengineonFPGA} shows the overview of the proposed stream computing. It consists of \textit{memory interfaces }to communicate with the system memory and \textit{computational pipelines}. There can be multiple pipelined chains in the FPGA that receive/send their data from/to memory through the ports available on the system (such as HP ports available on the Xilinx Zynq MPSoC). Each pipeline can consist of a few stages including \textit{read}, \textit{rearrange}, \textit{computation}, and \textit{write}. The \textit{read} stage fetches a stream of data from memory using the multiple wide-bit ports. The \textit{rearrange} stage reorganizes the data by splitting and concatenating operators to prepare the read data to be used in the successor stages. The \textit{computation} stage performs the main job in the given task.

A pipelined \textit{for} loop is usually used to implement each stage whose initiation interval (\textit{II}) defines its throughput. The \textit{II} of a pipelined loop is the minimum clock cycles between the starting point of the two consecutive loop iterations. If $ n $ and $ l $ denote the number of iterations in a loop and one iteration latency, respectively, then a pipelined loop requires $( n II+l )$ clock cycles to finish.  The stage with maximum \textit{II} restricts the total throughput and determines the execution time. If $ II_{max} $ and $ n_{max} $ denote the maximum \textit{II} and the maximum number of iterations of the stages in a pipeline, respectively, then the total clock cycles require to finish a pipeline is determined by Equ.~\ref{equ:exetime_pipeline}, where $l_{total}$ is the total latency of one iteration of all stages in the pipeline.
\begin{equation}
t_c=n_{max}II_{max}+l_{total}  \label{equ:exetime_pipeline}
\end{equation}

\section{Modelling Challenge}
\label{sec:ModellingChallenge}
Performance and power modeling are the key steps in designing a task efficiently on a heterogeneous system. Different types of modeling techniques have been proposed for GPUs on a system~\cite{7842980,6118939,7311199,7975297,6844867,6877247,8327055}. There are also a few articles proposing power and performance modeling for applications~\cite{6602306,8292049,6337536} on FPGA. Most of these approaches are application specific and consider the FPGA resource utilization or are simulation based. In contrast, we propose a high-level power and performance model suitable for an application implemented by HLS tools. 
This section addresses the power and performance modeling of streaming tasks running on an FPGA using the stream computing engine proposed in Fig.~\ref{fig:OverviewofStreamcomputingengineonFPGA}.

\subsection{Performance}
\label{subsec:Performance}
Traditionally, processing elements show their maximum performance if they can use their internal memory.  For example, utilizing different levels of cache memories in CPUs is the main factor of improving several application performances. GPUs utilize L2 cache memory along with device memories to improve the performance and provide parallel data access for many streaming processors in their architecture. FPGA also benefits from their internal BRAMs and distributed registers to save data temporarily during the computation. The FPGA internal memories have the capabilities to be used as cache memory tailored to the implemented task on the FPGA.

There have been many research activities on modifying the device and cache memory architectures to improve the performance on GPUs and CPUs, such that, repetitive applications with the data-reuse feature that can transfer the data once to the device or cache memories and benefits from their low latency and high-speed.
However, applications that require fresh data in each iteration, such as database processing, suffer from the high latency of accessing the system memory. Using zero-copy methodology and pipelining the data transfer with data computation are techniques to alleviate the relatively high latency of the system memory. The zero-copy technique maps the system memory as the device memory to be accessed directly by processing elements. The Nvidia Jetson TX1 can utilize the zero-copy using the unified memory programming technique, first introduced in CUDA 6.0. 

\begin{figure}
	\centering
	\includegraphics[width=0.8\linewidth]{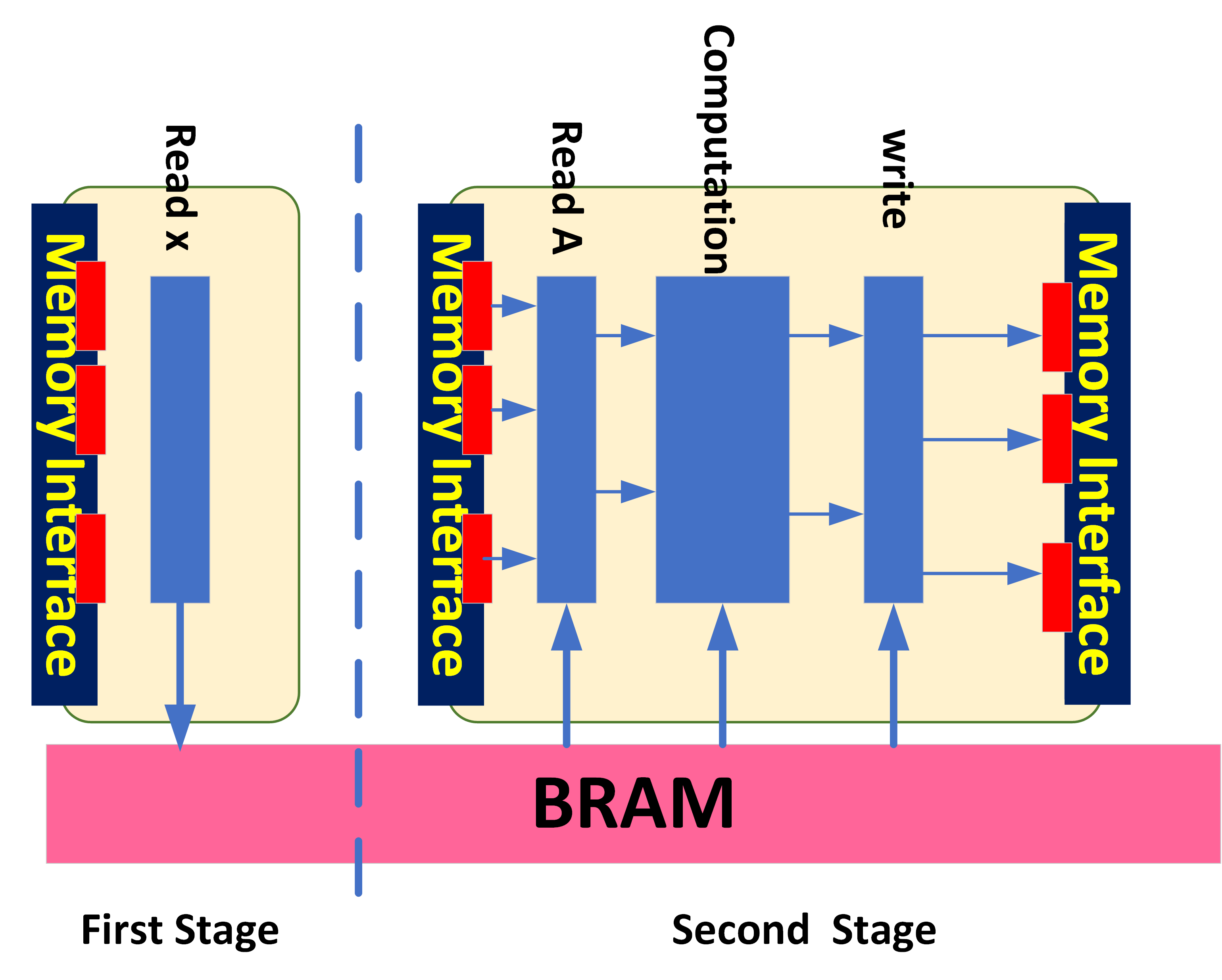}
	\caption{Dense matrix-vector multiplication: stream computing}
	\label{fig:Densematrixvectormultiplicationstreamcomputing}
\end{figure}

The proposed streaming engine in Fig.~\ref{fig:OverviewofStreamcomputingengineonFPGA} also benefits from the zero-copy technique to read data from the system memory which is pipelined with the computation. However, some part of a task may not be able to benefit from this technique. For example, in dense matrix-vector multiplication which is described by Equ.~\ref{equ:dense-matrix-vector-multiplication-1}, the vector $ x $ should be located in the FPGA internal memory (e.g., BRAM) to be reused for calculating each element of the output vector (i.e., $ y $).
In this case, a stream computing engine only with one stage (which is a pipelined \textit{for} loop) can transfer the $ x $ vector to the BRAM, then a streaming computing engine with three stages can read the elements of matrix $ A $ to generate the output. Fig.~\ref{fig:Densematrixvectormultiplicationstreamcomputing} shows this two-step stream processing. The first step is a pipelined loop with $ m $ iteration count, where $ m $ is the size of vector $ x $. The second step can be implemented by pipelined \textit{for} loops with $ n\times m $ iteration count, where $ n $ is the size of the output vector. Note that, both steps share the same memory interface, however, they are shown separately for the sake of clarity.
\begin{equation}
y=Ax \;\; where \;\; y_i = \sum_{j=0}^{j=m-1}{a_{i,j}x_j}\;\; i=0,1,...,n  \label{equ:dense-matrix-vector-multiplication-1} 
\end{equation}

\begin{figure}
	\centering
	\includegraphics[width=1\linewidth]{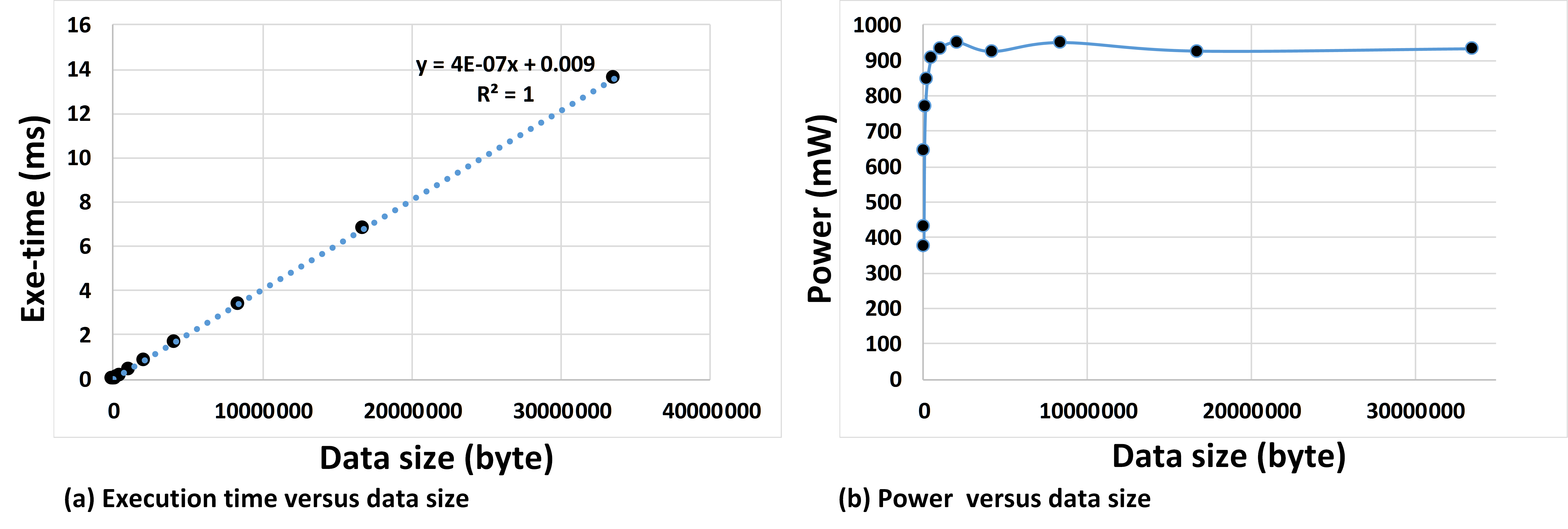}
	\caption{DeMV: performance and power versus data size}
	\label{fig:DeMVperformanceandpowerversusdatasize}
\end{figure}

The number of clock cycles for finishing each step in Fig.~\ref{fig:Densematrixvectormultiplicationstreamcomputing} can be described by Equ.~\ref{equ:exetime_pipeline}. The $ II $ on the first step is one as it uses burst data transfer and its loop iteration count is $ m $, therefore, it takes $ (m+l_1) $ clock cycles to finish, where $ l_1 $ is the latency of one iteration.
The initiation interval of the second step can be one (the optimize implementation is presented in Section~\ref{sec:ExperimentalSetups}) and its loop iteration count is $ n\times m $. Therefore, it takes $ (n\times m+l_2) $ clock cycle to finish, where $ l_2 $ is the latency of one ieration of all loops involved in the pipeline. Equ.~\ref{equ:mxv-clockcycles} represents the total clock cycles required to finish the whole task. If the size of input matrix is large enough to ignore the $ m $, $ l_1 $, and $ l_2 $ terms, the Equ.~\ref{equ:mxv-clockcycles-approximate} represents the performance of the task running on the FPGA which is directly defined by the data size (i.e., input matrix).
Fig.~\ref{fig:DeMVperformanceandpowerversusdatasize}(a) shows the execution time versus data size for the dense matrix vector multiplication.
\begin{equation}
T_c=\underbrace{(m+l_1)}_{Stage 1}+\underbrace{(n\times m+l_2)}_{Stage 2}  \label{equ:mxv-clockcycles}
\end{equation}

\begin{equation}
T_c\approx (n\times m)  \label{equ:mxv-clockcycles-approximate}
\end{equation}

Equation~\ref{equ:mxv-clockcycles} can be generalized to Equ.~\ref{equ:task-clockcycles-model} to model the performance of a task with $ S $ stages.
\begin{equation}
T_c= \sum_{s=0}^{S}{n_s\times II_s + l_s}  \label{equ:task-clockcycles-model}
\end{equation}

\subsection{Power and Energy}
\label{subsec:PowerandEnergy}
The power consumption of a task running on an accelerator usually consists of two main parts: the accelerator and the memory power consumptions.
The accelerator power consumption is defined based on the number of switching activities that happens in the underlying semiconductor fabrication cause by  value changes on the input data. In this section, we propose a simple model for the average FPGA and memory power for the stream computing engine proposed in the previous section. For the sake of simplicity, lets take the dense matrix-vector multiplication shown in Fig.~\ref{fig:Densematrixvectormultiplicationstreamcomputing}. If we assume that $ p_1 $ and $ p_2 $ represent the average power of the first and second stages, respectively, then Equ.~\ref{equ:mxv-task-average-power-1} or Equ.~\ref{equ:mxv-task-average-power-2} shows the total average power. Note that in this formula, we have ignored the iteration latencies (i.e., $ l_1 $ and $ l_2 $) in Equ.~\ref{equ:mxv-clockcycles}, for the sake of simplicity.

For large data sizes the second term in Equ.~\ref{equ:mxv-task-average-power-2} mainly defines the power and for small data sizes both terms are comparable and determine the total power. 
Fig.~\ref{fig:DeMVperformanceandpowerversusdatasize}(b) shows the power consumption versus data size for the dense matrix vector multiplication.
\begin{equation}
 P_{ave}= (mp_1+(n\times m)p_2)/(m + n\times m)  \label{equ:mxv-task-average-power-1}
\end{equation}

\begin{equation} 
  P_{ave} = \frac{m}{(m + n\times m)}p_1+\frac{n\times m}{m + n\times m}p_2 \label{equ:mxv-task-average-power-2}
\end{equation}

This formula can be generalized for tasks with more stages as Equ.~\ref{equ:task-average-power-model} where $ S $ in the number of stages and $ p_s $ and $ n_s $ represent the power and data size of each stage.
\begin{equation} 
P_{ave} = \sum_{s=0}^{S}{\frac{n_s}{(\sum_{i=0}^{S}{n_i})}p_s} \label{equ:task-average-power-model}
\end{equation}

%Fig.~\ref{fig:Histogrampowerversusdatasize} shows the average FPGA and memory power of the histogram. As its implementation consists of two parts, its power model is similar to Equ.~\ref{equ:mxv-task-average-power-2}. Therefore, for large data sizes we can see a steady state situation in both memory and FPGA power, as one of the part become dominant.  

\section{Scheduling Challenge}
\label{sec:SchedulingChallenge}
Task scheduling among multiple processors in a system is a mature subject with extensive research activities. However, they need a kind of modification and tuning to be applied to a new system such as the heterogeneous FPGA+GPU embedded system considered in this paper. For the sake of simplicity, we only consider the scheduling problem in data parallel tasks. In this case, we should divide the data between the FPGA and GPU to achieve high performance. For this purpose, both FPGA and GPU should utilize their maximum performance and should finish their tasks at the same time. In other words, a load balancing is required for maximum performance. 
Here we only propose a simple task division between FPGA and GPU for large data sizes so that the behavior of the system is more predictable and depends on the data sizes. Considering this assumption, the FPGA and GPU execution times are directly proportional to the data size which are shown in Equs.~\ref{equ:fpga-per-1} and~\ref{equ:gpu-per-1}, where $ n_{fpga} $ and $ n_{gpu} $ are the data sizes on the FPGA and GPU, respectively, $ a $ and $ b $ are constant that can be determined by data patterns.
In this case, task division and load balancing can be described by Equ.~\ref{equ:task-division} and~\ref{equ:load-balancing}, respectively. Solving these equations results in Equ~\ref{equ:result-1}. If $ \alpha $ represents the GPU speed-up compared to the FPGA (i.e., $ \alpha=a/b $), then Equ.~\ref{equ:fpga-per-6} shows the task scheduling solution. Section~\ref{sec:ExperimentalSetups} empirically evaluates this task scheduling solution.

\begin{equation} 
t_{fpga}=a.n_{fpga}  \label{equ:fpga-per-1}
\end{equation}
\begin{equation} 
t_{gpu}=b.n_{gpu}\label{equ:gpu-per-1}
\end{equation}

\begin{equation} 
n_{fpga}+n_{gpu} = n \label{equ:task-division}
\end{equation}
 
\begin{equation} 
a.n_{fpga}=b.n_{gpu} \label{equ:load-balancing}
\end{equation}

\begin{equation} 
n_{fpga} = \frac{b}{a+b}n\;\; and\;\; n_{gpu} = \frac{a}{a+b}n \label{equ:result-1}
\end{equation}

\begin{equation} 
n_{fpga} = \frac{1}{\alpha+1}n\;\; and\;\; n_{gpu} = \frac{\alpha}{1+\alpha}n \label{equ:fpga-per-6}
\end{equation}

\section{Experimental Setups}
\label{sec:ExperimentalSetups}

One idea to have a research platform that combines FPGA and GPU is to connect the Xilinx Zynq MPSoC and Jetson TX1. However, since suitable systems deivers are not available from the SoC vendors, we decided to connect the Xilinx Virtex-7 FPGA to the Jetson TX1 board. Table~\ref{tbl:ZynqMPSoConZCU102boardversusVirtex7onVC707board} compares the Virtex-7 FPGA feature with that of the Zynq MPSoC and it is shown that the two FPGAs are very close in terms of the available resources. The experimental results also show the low power consumption of the Virtex-7. 

\begin{table}
	\caption{ZynqMPSoC on ZCU102 board versus Virtex 7 on VC707 board}
	\label{tbl:ZynqMPSoConZCU102boardversusVirtex7onVC707board}
	\includegraphics[width=0.8\linewidth]{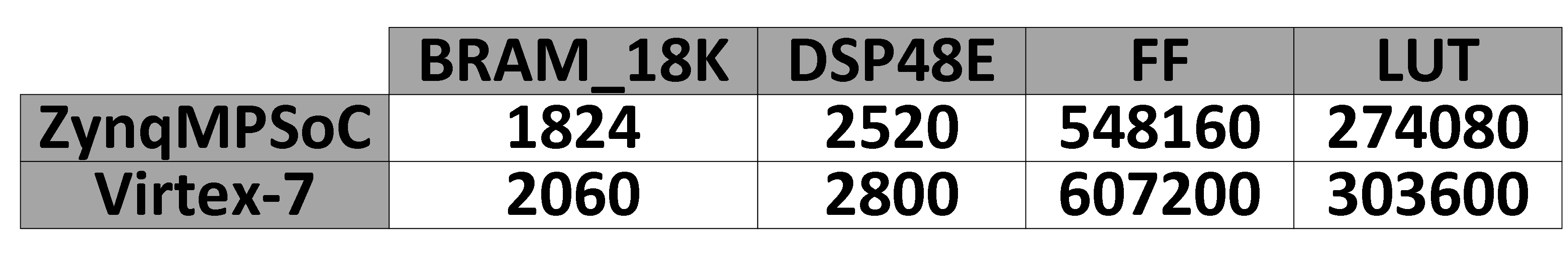}	
\end{table}

Although the FPGA is connected to the Jetson TX1 over a PCIe bus, it still can be used to study some of the features and behaviors of the heterogeneous embedded systems if we assume that the input data is available in the FPGA onboard memory to which FPGA has direct access with a 512-bit wide AXI bus.  

\begin{figure*}
 	\centering
 	\includegraphics[width=0.8\linewidth]{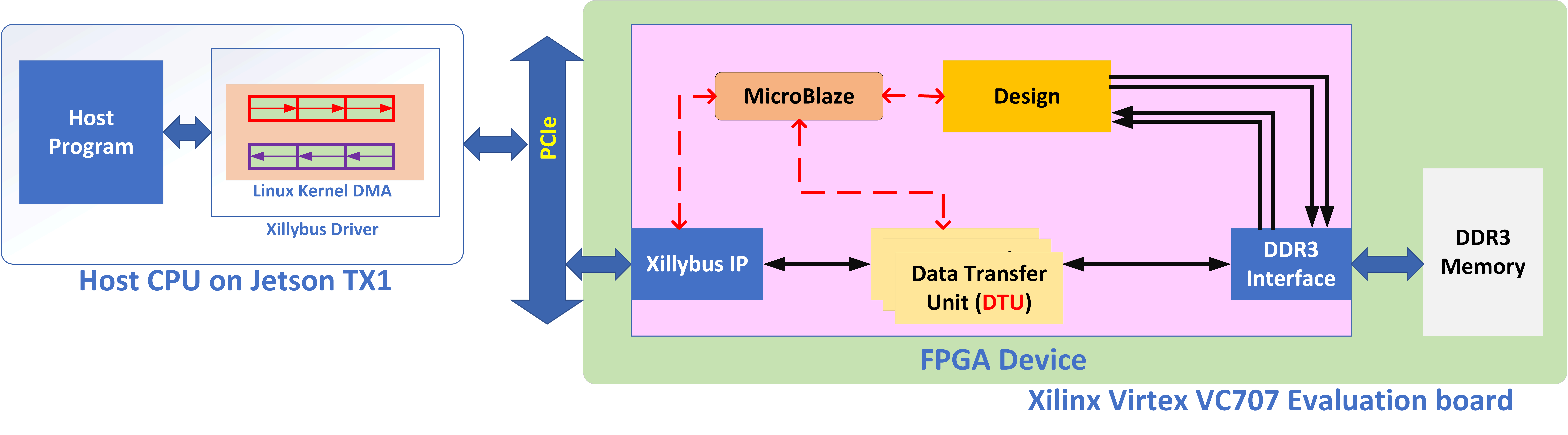}
 	\caption{The experimental FPGA+GPU architecture }
 	\label{fig:TheFPGAarchitecture}
 \end{figure*}

Fig.~\ref{fig:TheFPGAarchitecture} illustrates the system hardware architecture through which the FPGA is connected to the Jetson TX1 board over a 4x PCIe bus. The FPGA hardware is comprised of two sections. 

The first section, consisting of the Xillybus IP~\cite{Xillybus-ref}, data transfer unit (DTU) and DDR3 interface, provides the data path between the PCIe and the onboard DDR memory. The Xillybus IP provides a streaming data transfer over PCIe, DTU receives this stream and copies that into the DDR3 memory using master AXI bus through DDR3 interface. Fig. 7 shows the high-level C code for the write-to-memory parts of the data transfer unit (DTU) synthesizable with the Xilinx Vivado HLS. It consists of a pipelined loop that receives a unit of data and writes it to the memory in each clock cycle. The maximum memory bandwidth provided by the first path is 800MBytes/s mainly because of the PCIe Gen1 used in the Jetson TX1 which is compatible with the Xilinx IP core located in the Vitex-7 FPGA.

\begin{figure}
	\centering
	\includegraphics[width=0.8\linewidth]{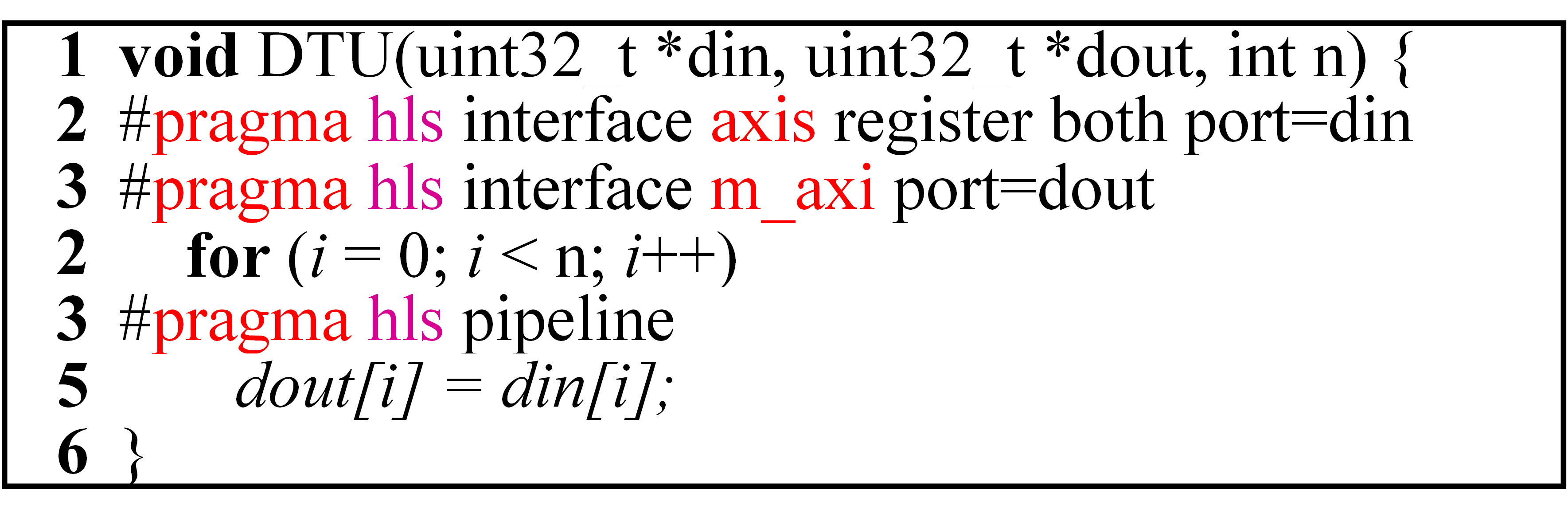}
	\caption{Data transfer unit (DTU)}
	\label{fig:DatatransferunitDTU}
\end{figure}
  
The second path consists of the user design and the DDR3 interface which can provide up to $ 6.4GByte/s $ using a 512-bit wide-bus at the frequency of 100MHz. 

A Xilinx MicroBlaze software core is used to control the activation of different paths in the FPGA. For this purpose, it runs a firmware that receives different commands from the host processor on the Jetson TX1 through PCIe and activates the application design. This controller also informs the system when the task execution finishes. Onboard memory management and allocation are other tasks of the controller. In summary, the firmware running on the MicroBlaze performs the following functions: 
\begin{itemize}
	\item \textbf{initFpga}: This function performs the FPGA initialization and prepares the memory allocation tables on the MicroBlaze.
	\item 	\textbf{fpgaMalloc}: This function gets two arguments: the ID variable and the size of the memory requested for allocation. It returns the start address of the allocated memory or -1 in the case of failure of the memory allocation process.
	\item \textbf{startAccel}: Receiving this command, the MicroBlaze activates the design to perform its task.
	\item \textbf{fpgaFree}: This function resets the memory allocation table corresponding to the allocated memories.
\end{itemize}

The algorithm under acceleration is described in HLS C/C++ that is synthesizable by the Xilinx Vivado-HLS which uses the AXI master protocol to send/receive data to/from DDR3 memory using the burst data transfer protocol. 

\section{Experimental Results}
\label{sec:ExperimentalResults}
Three different tasks are studied as our benchmarks in this section to evaluate the potential of embedded FPGA+GPU system in providing a high-performance and low energy consumption system. The results show that the concurrent execution between FPGAs and GPUs can result in 2x performance or energy reduction after efficient algorithm implementation, correct workload balancing and data transfer optimizations. These three algorithms are: \textit{histogram}, \textit{dense matrix-vector multiplication} (DeMV), and \textit{sparse matrix-vector multiplication} (SpMV). The experimental setup explained in Section~\ref{sec:ExperimentalSetups} is used for real measurements. In addition, for the sake of completeness, the two distinct Jetson TX1 and ZynqMpsoC systems are also used to generate results for comparison even if they are not connected. 

\subsection{Histogram}
\label{subsec:histogram}
\begin{figure}
	\centering
	\includegraphics[width=0.8\linewidth]{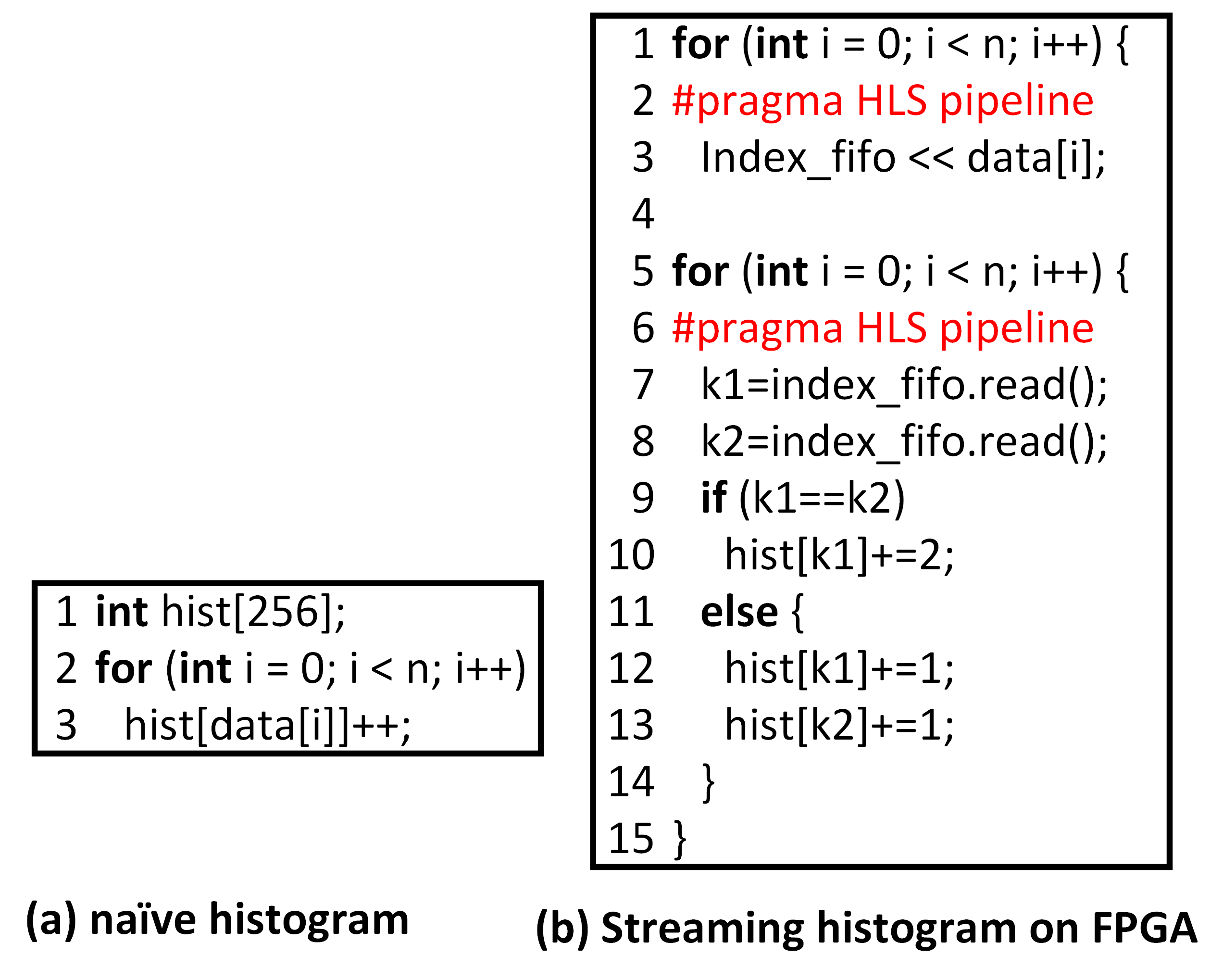}
	\caption{Histogram algorithm}
	\label{fig:Histogramalgorithm}
\end{figure}

Fig.~\ref{fig:Histogramalgorithm}(a) shows the original histogram pseudo code. It consists of a \textit{for} loop iterating over the entire input data, modifying the \textit{hist} array (as the histogram bin holder) using the input data as the index to access the corresponding bin. This na\"ive algorithm can be easily pipelined on the FPGA using the Xilinx Vivado-HLS tool, however because of the data dependency between two consecutive loop iterations (note that two consecutive iterations can modify the same bin in the \textit{hist} array), the obtained initiation interval is 2 which reduces the performance. Fig.~\ref{fig:Histogramalgorithm}(b) shows one hardware thread of the stream computing implementation of the histogram suitable for FPGA. It consists of two stages. The first stage from Line 1 to Line 3 reads data from the memory using the burst protocol i.e., reading a data per clock cycle or \textit{II=1}. The second stage modifies the bins. As the initiation interval of the pipelined loop for the \textit{hist} modification is 2, this loop reads two data and modifies the \textit{hist} by resolving the potential conflict using the \textit{if} condition at Line 9. As this stage reads two data values in each iteration and its $ II=2 $, then the average number of data read per clock cycle is $ 2/2=1 $, that means, it consumes the data at the same pace that is generated by the first stage.
As the total memory bus width in ZynqMPSoC and Virtex 7 is 512 and if each pixel in the input image is represented by an 8-bit code, then $ 512/8=64 $ hardware threads can be instantiated to implement histogram on the FPGA.

\begin{figure*}
	\centering
	\includegraphics[width=1\linewidth]{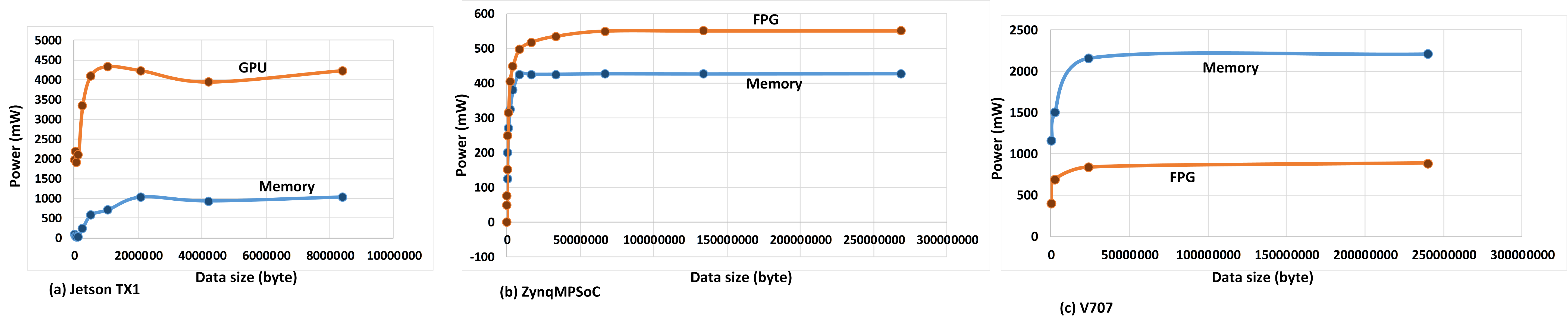}
	\caption{Histogram power consumption}
	\label{fig:Histogrampowerconsumption}
\end{figure*}

\begin{figure}
	\centering
	\includegraphics[width=1\linewidth]{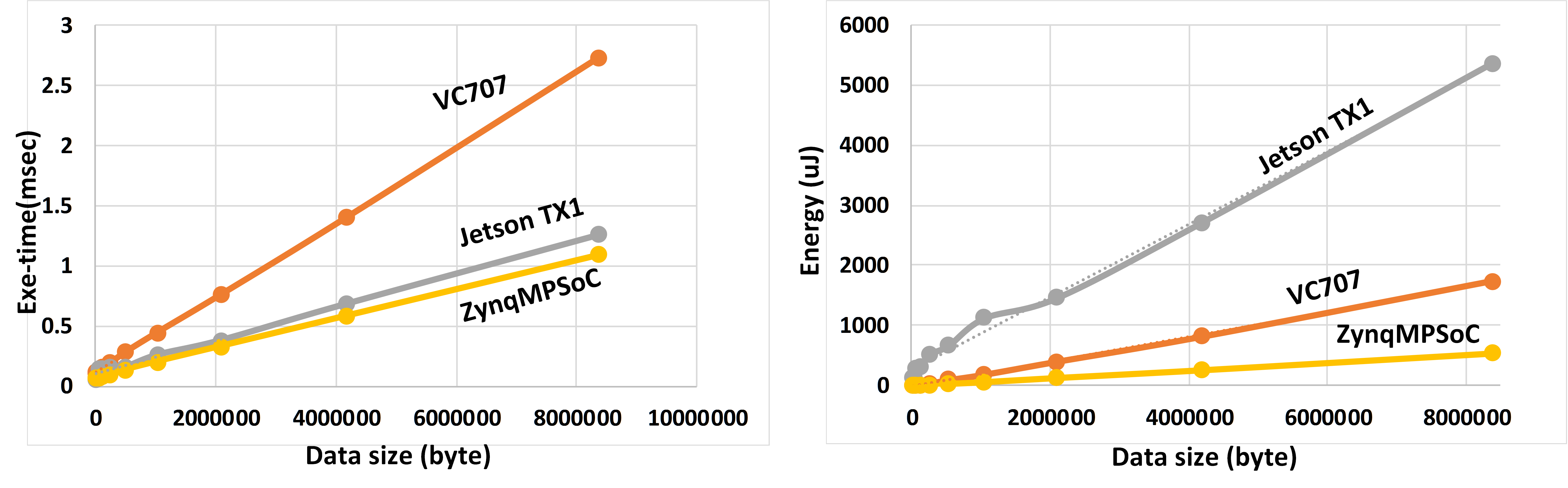}
	\caption{Histogram execution time}
	\label{fig:Histogramexecutiontime}
\end{figure}

Table~\ref{tbl:HistogramFPGAresourceutilization} shows the resource utilization of 64-thread implementations of the histogram on Zynq MPSoC and Vertex7 FPGAs. The power consumptions of histogram task versus the data sizes on the three platforms are shown in Fig.~\ref{fig:Histogrampowerconsumption}. As mentioned in Subsection~\ref{subsec:PowerandEnergy}, the power consumption consists of two components: the accelerator (i.e., GPU or FPGA) and the memory. As can be seen from these diagrams, running the histogram on the zynq MPSoC consumes the least power among the three platforms. As the two Jetson TX1 and Zynq MPSoC utilize embedded memories, their memory power consumption is less than the Virtex 7 memory power requirement. The GPU consumes about $ 7.7 $ and $ 4.8 $ times more power than Zynq MPSoC and Virtex 7. Fig.~\ref{fig:Histogramexecutiontime} compares the histogram execution time and energy consumption versus the data size, considering the three platforms. As can be seen, although the performance of this task is very close on the Jetson TX1 and Zynq MPSoC, its energy consumption on the Zynq MPSoC is about 10 times less than that of the Jetson TX1.

\begin{table}
	\caption{Histogram FPGA resource utilization}
	\label{tbl:HistogramFPGAresourceutilization}
	\includegraphics[width=1\linewidth]{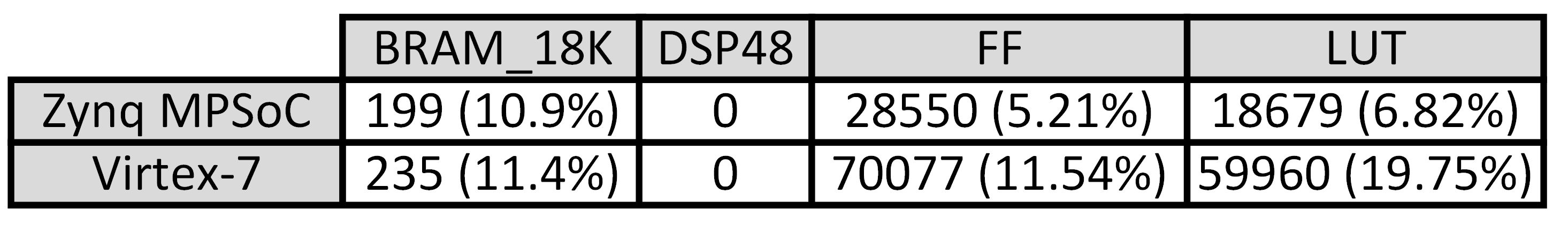}	
\end{table}

According to the performance diagram of Fig.~\ref{fig:Histogramexecutiontime}, the speed-up factors (i.e., $ \alpha $ in Equ.~\ref{equ:fpga-per-6}) for the Jetson to the Zynq MPSoC and Virtex 7 FPGAs are $ 0.85 $ and $ 2.0 $ for large data sizes. Table~\ref{tbl:HistogramFPGAJetsonscheduling} shows the results of task division between the GPU and FPGA using Equ.~\ref{equ:fpga-per-6}. to divide an input data size of $ 8388608 bytes $ between the GPU and FPGA. The table shows $ 1.79 $ and $ 2.29 $ times improvement in performance and energy consumption, respectively, if the task is divided between the Zynq and Jetson compared to only GPU running the application. In addition, it shows $ 1.18 $ and $ 1.45 $ times improvement in performance and reduction in energy consumption, respectively, if the task is divided between the Virtex 7 and Jetson compared to only the GPU running the application.

\begin{table}
	\caption{Histogram FPGA\&Jetson task division for $ 8388608 bytes $ of data}
	\label{tbl:HistogramFPGAJetsonscheduling}
	\includegraphics[width=1\linewidth]{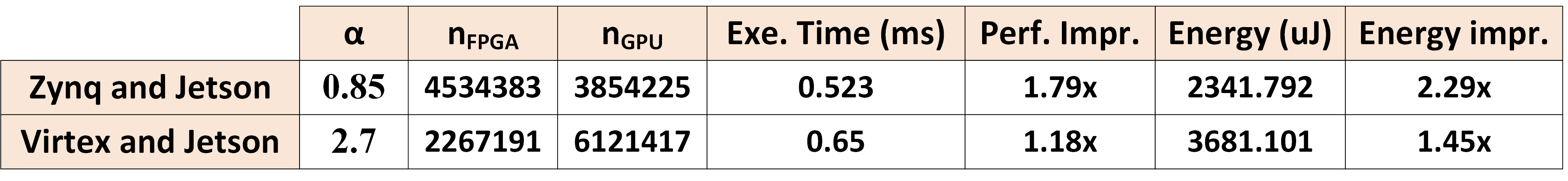}
\end{table}

\subsection{Dense Matrix-Vector Multiplication (DeMV)}
\label{subsec:DenseMatrixVectorMultiplication}
\begin{figure}
	\centering
	\includegraphics[width=0.8\linewidth]{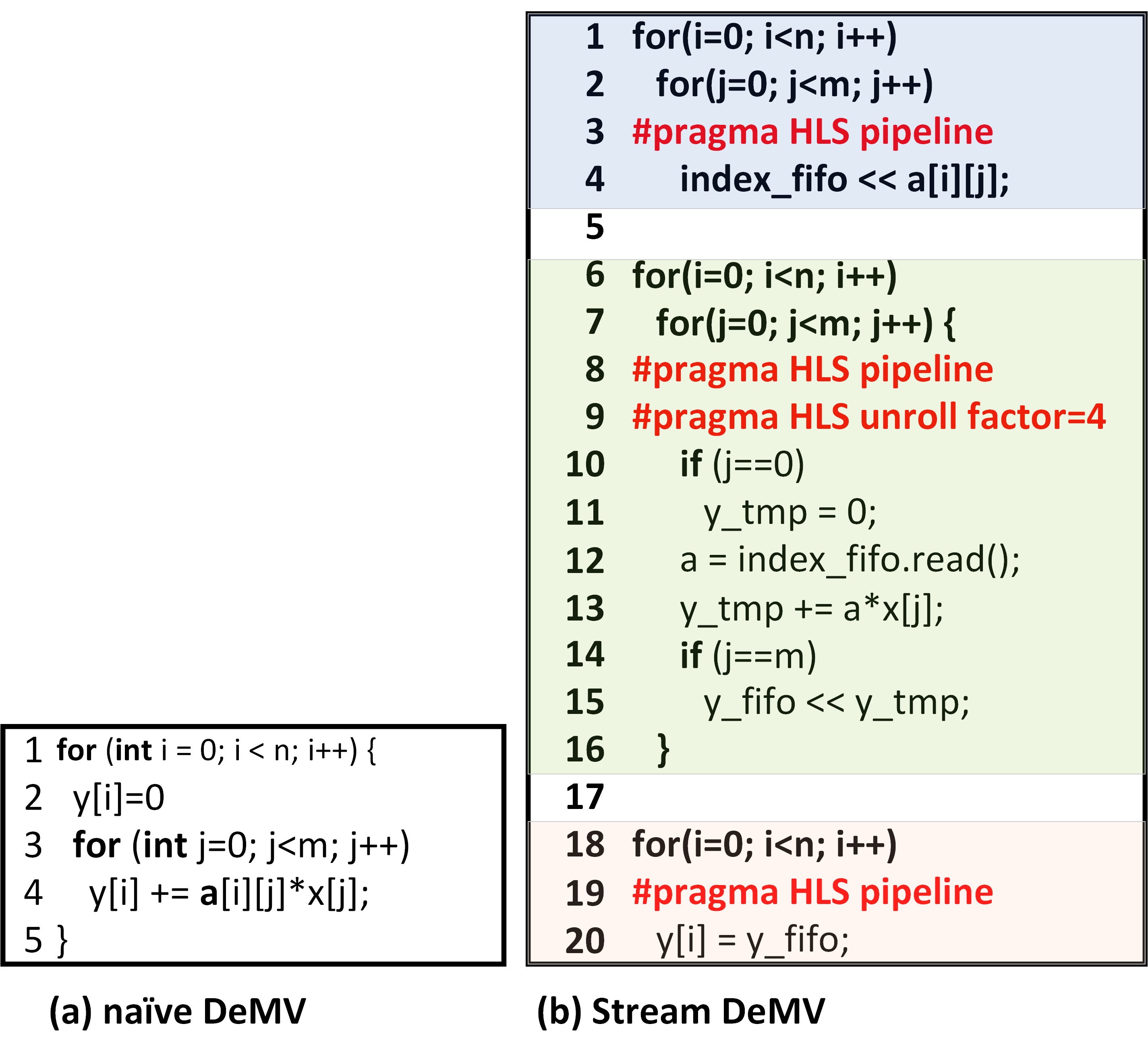}
	\caption{DeMV Pseudo-Codes}
	\label{fig:DeMVPseudoCodes}
\end{figure}

Fig.~\ref{fig:DeMVPseudoCodes}(a) shows the na\"ive pseudo-code for the dense matrix-vector multiplication which consists of two nested loops performing the accumulation statement at Line 4. Fig.~\ref{fig:DeMVPseudoCodes}(b) shows one thread of the pipelined version of this task which consists of two stages. The first stage from Line 1 to Line 4 reads the data on each clock cycle. The pipelined loop in the second stage from Line 6 to Line 12 shows an \textit{II=4} after synthesis which reduces the total performance. In order to address this issue, we have unrolled this loop with a factor of 4 to read four data values in each iteration. Therefore, it consumes the data at the same pace that is generated by the first stage. This results in the \textit{II=1} for the whole design. Table~\ref{tbl:DeMVFPGAresourceutilization} shows the FPGA resource utilization. 

\begin{figure*}
	\centering
	\includegraphics[width=1\linewidth]{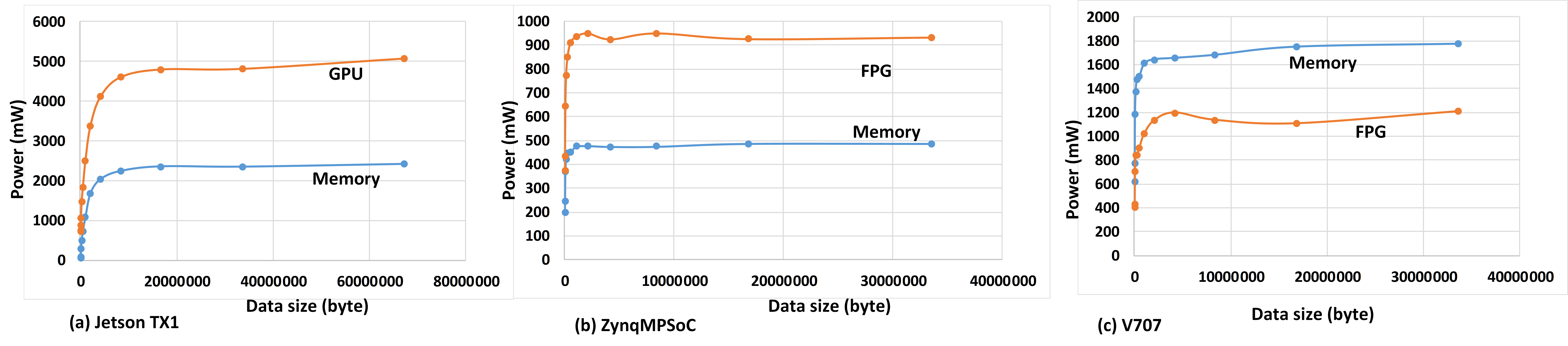}
	\caption{DeMV power consumption}
	\label{fig:DeMVpowerconsumption}
\end{figure*}

\begin{figure}
	\centering
	\includegraphics[width=1\linewidth]{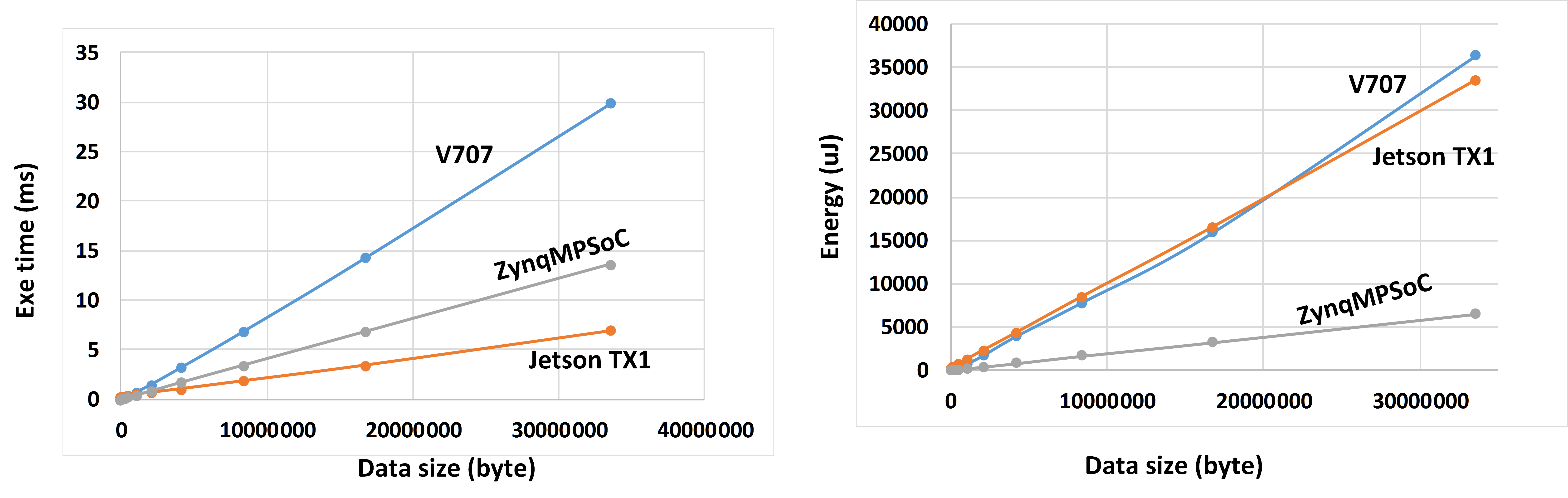}
	\caption{DeMV performance and energy consumption}
	\label{fig:DeMVPerfroamnceconsumption}
\end{figure}

Fig.~\ref{fig:DeMVpowerconsumption} shows the power consumption diagrams of running DeMV on the three embedded platforms. The GPU consumes up to 5.20 and 4.3 times more power than Zynq MPSoC and Virtex 7 FPGAs. Fig.~\ref{fig:DeMVPerfroamnceconsumption} compares the DeMV performance and energy consumptions. Similar to the histogram task, the Zynq shows much less energy consumption compared to the other PEs. 

\begin{table}
	\caption{DeMV FPGA resource utilization}
	\label{tbl:DeMVFPGAresourceutilization}
	\includegraphics[width=1\linewidth]{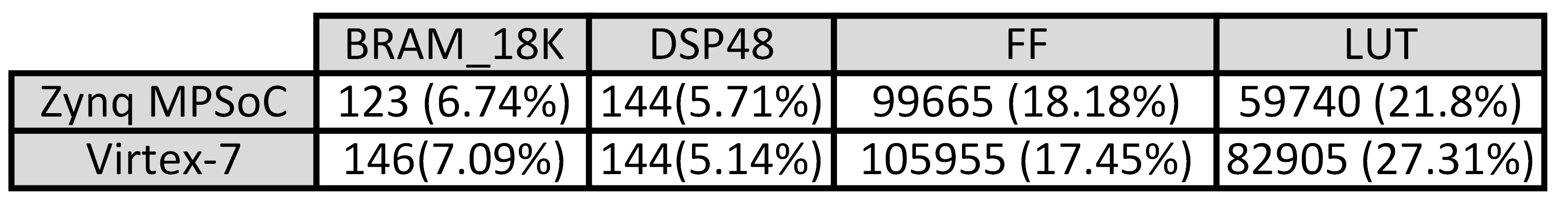}
\end{table}

According to the performance diagram of Fig.~\ref{fig:DeMVPerfroamnceconsumption}, the speed-up factors (i.e., $ \alpha $ in Equ.~\ref{equ:fpga-per-6}) for the Jetson to the Zynq MPSoC and Virtex 7 FPGAs are $ 0.51 $ and $ 0.23 $ for large data sizes. Table~\ref{tbl:DeMVFPGAJetsonscheduling} shows the results of task division between the GPU and FPGA using Equ.~\ref{equ:fpga-per-6} to divide an input data size of $ 33554432 $ between the GPU and FPGA. The table shows $ 1.48 $ and $ 1.19 $ times improvement in performance and energy consumption, respectively, if the task is divided between the Zynq and Jetson compared to only GPU running the application. In addition, it shows $ 1.22\times $ improvement in performance and slightly increase (i.e., $ 1-0.96=0.04\times $) in energy consumption, respectively, if the task is divided between the Virtex 7 and Jetson compared to only the GPU running the application. 

\begin{table}
	\caption{DeMV FPGA\&Jetson task division for data size of $ 33554432 $ }
	\label{tbl:DeMVFPGAJetsonscheduling}
	\includegraphics[width=1\linewidth]{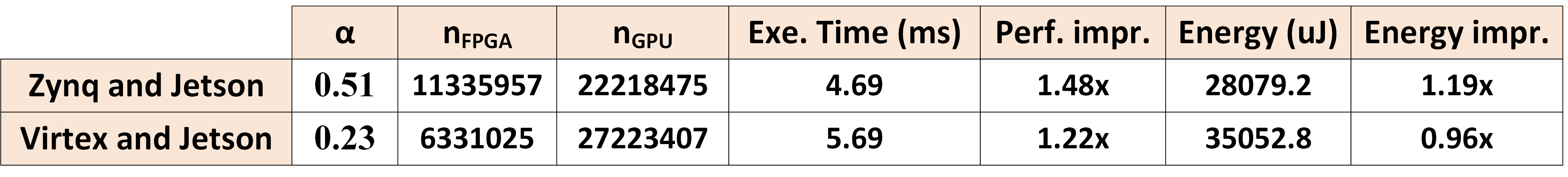}
\end{table}

\subsection{Sparse Matrix-Vector Multiplication (SpMV)}
\label{subsec:SparseMatrixVectorMultiplication}
\begin{figure}
	\centering
	\includegraphics[width=0.9\linewidth]{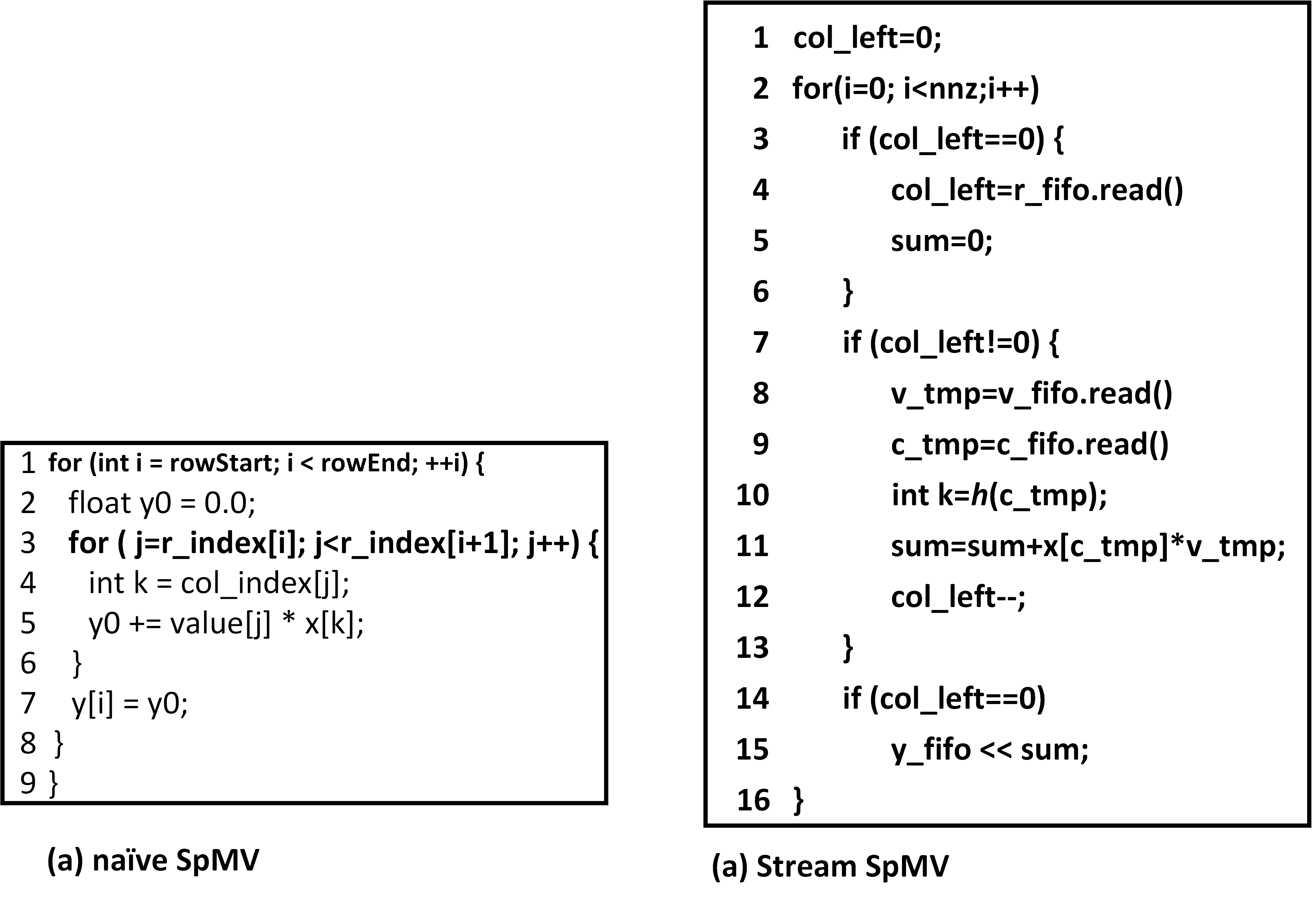}
	\caption{SpMV Pseudo-Codes}
	\label{fig:SpMVPseudoCodes}
\end{figure}

The pseudo-code of the sparse matrix-vector multiplication, based on the Compressed Sparse Row (CSR) representation~\cite{6375570}, is shown in Fig.~\ref{fig:SpMVPseudoCodes}(a). One thread of the corresponding streaming computation suitable for the FPGA is shown in Fig.~\ref{fig:SpMVPseudoCodes}(b). The table is shown in Fig.~\ref{tbl:SpMVFPGAresourceutilization} contains the FPGA resource utilization of this task after synthesis.

\begin{figure*}
	\centering
	\includegraphics[width=1\linewidth]{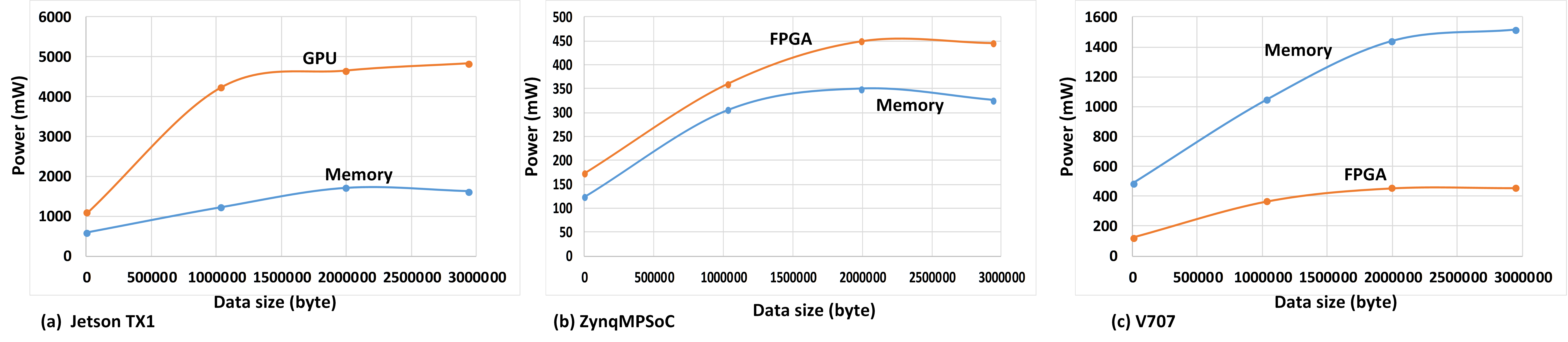}
	\caption{SpMV power consumption}
	\label{fig:SpMVpowerconsumption}
\end{figure*}

\begin{figure}
	\centering
	\includegraphics[width=1\linewidth]{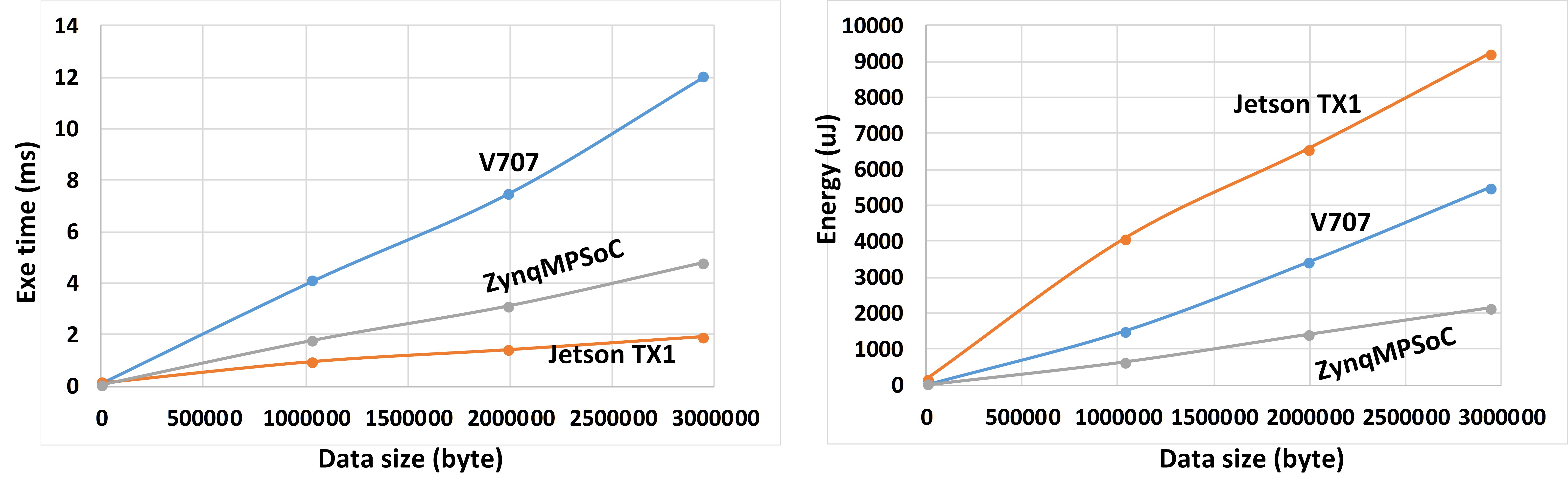}
	\caption{SpMV performance and energy consumption}
	\label{fig:SpMVperformanceandenergyconsumption}
\end{figure}

The SpMV power consumptions versus data sizes for the three platforms are shown in Fig.~\ref{fig:SpMVpowerconsumption}. As can be seen, the Zynq MPSoC consumes the least power compared to the other platforms. Fig.~\ref{fig:SpMVperformanceandenergyconsumption} compares the performance and energy consumption of the SpMV on Jetson TX1, Zynq MPSoC and Virtex 7. 

\begin{table}
	\caption{SpMV FPGA resource utilization}
	\label{tbl:SpMVFPGAresourceutilization}
	\includegraphics[width=1\linewidth]{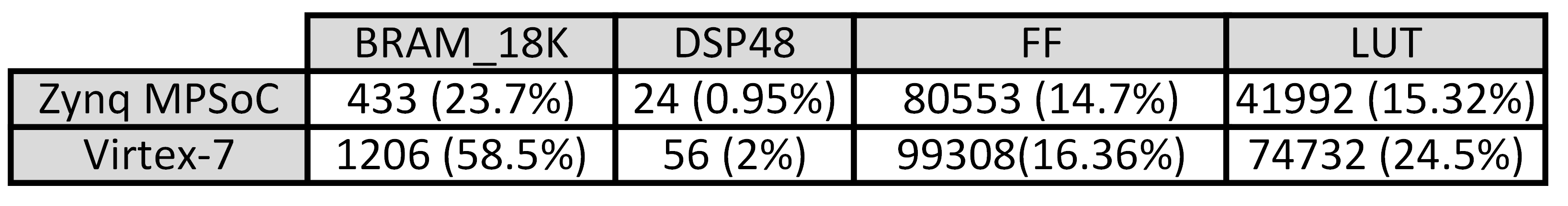}
\end{table}

According to the performance diagram of Fig.~\ref{fig:SpMVperformanceandenergyconsumption}, the speed-up factors (i.e., $ \alpha $ in Equ.~\ref{equ:fpga-per-6}) for the Jetson to the Zynq MPSoC and Virtex 7 FPGAs are $ 3.2 $ and $ 6.4 $ for large data sizes. Table~\ref{tbl:SpMVFPGAJetsonscheduling} shows the results of task division between the GPU and FPGA using Equ.~\ref{equ:fpga-per-6} to divide an input data size of $ 2943887 $ between the GPU and FPGA. The table shows $ 1.46 $ and $ 1.23 $ times improvement in performance and energy consumption, respectively, if the task is divided between the Zynq and Jetson compared to the only GPU running the application. In addition, it shows $ 1.15\times $ and $ 1.1\times $ improvement in performance and reduction in energy consumption, respectively, if the task is divided between the Virtex 7 and Jetson compared to the only GPU running the application.

\begin{table}
	\caption{SpMV FPGA\&Jetson task division for data size of $ 2943887  $}
	\label{tbl:SpMVFPGAJetsonscheduling}
	\includegraphics[width=1\linewidth]{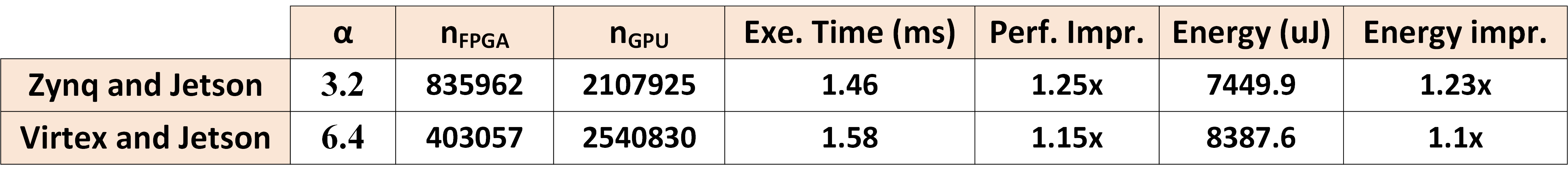}
\end{table}

\section{Conclusions}
\label{sec:Conclusions}
This paper has studied the challenges and opportunities that designers will face when using a heterogeneous embedded FPGA+GPU platform. The challenges are categorized into three groups: design, modeling, and scheduling. Using the image histogram operation, the paper has clarified the trade-off between performance and energy consumption by distributing the task between the GPU and FPGA. Focusing on the FPGA, then the paper has proposed a stream computing engine with the corresponding modeling technique to cope with the design and modeling challenges, respectively. A scheduling technique has been proposed to improve the performance and energy consumption by distributing a parallel task between the FPGA and GPU. Three applications including histogram, dense matrix-vector multiplication, and sparse matrix-vector multiplication are used to evaluate the proposed techniques. The experimental results have shown improvement in performance and reduction in energy consumption by factors of $ 1.79\times $ and $ 2.29\times $, respectively.

\bibliographystyle{ACM-Reference-Format}
\bibliography{hip3es}

\end{document}